\documentclass[12pt,a4paper]{article}
\usepackage[left=1in, right=1in, top=1in, bottom=1in]{geometry}

\usepackage[utf8]{inputenc}

\usepackage{amsmath}
\usepackage{amsfonts}
\usepackage{amssymb}

\usepackage{pgffor}
\usepackage{amssymb,times,graphicx, subfigure}
\usepackage[english]{babel}
\usepackage[varg]{txfonts}
\usepackage{xspace}
\usepackage{longtable,lscape,supertabular}
\usepackage{rotating}

\usepackage[table]{xcolor}
\usepackage{multicol}

\usepackage{hyperref}
\hypersetup{
  colorlinks=true,
  linkcolor=blue,
  citecolor=black,
  filecolor=magenta,
  urlcolor=blue,
  pdfpagemode=UseThumbs,
  pdfstartview=FitH,
  pdfauthor={Elvire De Beck},
  pdfsubject={2020 Astro Decadal},
  pdfkeywords={Stellar physics, Stellar Evolution, Outflows, Chemistry}
}

\usepackage{color}

\usepackage{titlesec} 
\titleformat{\section}  [runin]				{\normalfont\large\bfseries}{\thesection.}{0.25em}{}
\titleformat{\subsection}[runin]		 	{\normalfont\large\bfseries}{\thesubsection.}{0.25em}{}
\titleformat{\subsubsection}[runin]  	{\normalfont\normalsize\bfseries}{\thesubsubsection.}{1em}{}
\titlespacing*{\section}   		{0pt}{0.25\baselineskip}{0\baselineskip}
\titlespacing*{\subsection}   {0pt}{0.25\baselineskip}{0.5\baselineskip}

\usepackage[square,numbers,sort&compress,super]{natbib}
\usepackage{xspace}
\def\irc{IRC\,+10\,216\xspace}

\def\cit6{CIT~6\xspace}

\def\rstar{$R_{\star}$\xspace}
\def\msun{$M_{\odot}$\xspace}
\def\mzams{$M_{\rm ZAMS}$\xspace}

\def\mytitle{The fundamentals of outflows from evolved stars}

\usepackage{fancyhdr}
\usepackage[inline]{enumitem}
\newlist{todolist}{itemize}{2}
\setlist[todolist]{label=$\square$}
\usepackage{pifont}
\newcommand{\cmark}{\ding{51}}%
\newcommand{\done}{\color{blue}\rlap{$\square$}{\raisebox{2pt}{\large\hspace{1pt}\cmark}}%
\hspace{-2.5pt}}
\usepackage{tabto}

\def\arxivprefixesep{:}

\newcommand{\eprint}[2][]{%
{\tt\if!#1!#2\else#1\arxivprefixesep\ignorespaces#2\fi}%
}

\usepackage[strict]{changepage}
\usepackage{framed}
\definecolor{formalshade}{rgb}{0.95,0.95,1}

\usepackage{tcolorbox}
\tcbuselibrary{breakable}
 {\end{tcolorbox}}

 \def\hrulefill{\leavevmode\leaders\hrule height 6pt\hfill\kern 0pt}

\newcolumntype{P}[1]{>{\raggedright\arraybackslash}p{#1}}

\usepackage[font=small,labelfont=sc,bf]{caption}

\begin{document}

\setcounter{page}{0}

\thispagestyle{empty}
\begin{center}
{\Large \bf \mytitle}\\
An Astro2020 Science White Paper\\
{\color{orange!50!white}\noindent\hrulefill}
\end{center}
%\vspace{3cm}
%\begin{todolist*}
\begin{enumerate*}[label=$\square$,font=\color{black},itemjoin={\tab}]

\item[\done] {\color{blue}Stars and Stellar Evolution  }\\
\item[\done] {\color{black} Resolved Stellar Populations and their Environments}
\item Galaxy Evolution       
\item Star and Planet Formation
\item  Planetary Systems     
\item Formation and Evolution of Compact Objects  
\item Cosmology and Fundamental Physics
\item Multi-Messenger Astronomy and Astrophysics 
\end{enumerate*}
% \end{todolist*}

\vspace{0.5cm}
\noindent\textbf{\underline{Principal author: }}
\hspace{0.3cm} \textbf{Elvire De Beck}  \\\\[-1ex]
Division of Astronomy and Plasma Physics, \\
Department of Space, Earth and Environment, \\
Chalmers University of Technology, Sweden\\
\texttt{elvire.debeck[at]chalmers.se};  +46317725545 \\[-2ex]

\noindent\textbf{\underline{Co-authors: }} \\\\[-1ex]
\textbf{M. L. Boyer}, Space Telescope Science Institute, USA;
\textbf{V. Bujarrabal}, Observatorio Astronomico Nacional (OAN, IGN), Spain;
\textbf{L. Decin}, KU Leuven, Belgium; 
\textbf{J. P. Fonfr{\'i}a},	Instituto de F{\'i}sica Fundamental (IFF), Consejo Superior de Investigaciones Cient{\'i}ficas (CSIC), Spain;
\textbf{M. Groenewegen}, Royal Observatory Belgium, Belgium; 
\textbf{S. H{\"o}fner}, Department of Physics and Astronomy,  University of Uppsala, Sweden;
\textbf{O. Jones}, UK Astronomy Technology Centre, UK;
\textbf{T. Kami{\'n}ski}, Center for Astrophysics, Harvard \& Smithsonian, Cambridge, USA;
\textbf{M. Maercker}, Department of Space, Earth and Environment, Chalmers University of Technology, Sweden;
\textbf{P. Marigo}, Department of Physics and Astronomy, University of Padova, Italy;
\textbf{M. Matsuura}, School of Physics and Astronomy, Cardiff University, Queen's Buildings, UK;
\textbf{M. Meixner}, Space Telescope Science Institute, USA;
\textbf{G. Quintana Lacaci Mart{\'i}nez}, Instituto de F{\'i}sica Fundamental (IFF), Consejo Superior de Investigaciones Cient{\'i}ficas (CSIC), Spain;
\textbf{P. Scicluna}, Academia Sinica Institute of Astronomy and Astrophysics, Taiwan;
\textbf{R. Szczerba}, Nicolaus Copernicus Astronomical Center, Poland; 
\textbf{L. Velilla Prieto}, Department of Space, Earth and Environment, Chalmers University of Technology, Sweden;
\textbf{W. Vlemmings}, Department of Space, Earth and Environment, Chalmers University of Technology, Sweden;
\textbf{M. Wiedner}, Sorbonne University, Observatoire de Paris, CNRS, LERMA, France

\vspace{0.5cm}

\begin{tcolorbox}[breakable,colback=orange!5!white,colframe=orange]
%\begin{center}
{\underline{\textbf{Abstract:}}}
Models of the chemical evolution of the interstellar medium, galaxies, and the Universe rely on our understanding of the amounts and chemical composition of the material returned by stars and supernovae. Stellar yields are obtained from stellar-evolution models, which currently lack predictive prescriptions of stellar mass loss, although it significantly affects stellar lifetimes, nucleosynthesis, and chemical ejecta.
%Mass loss can dominate the course of stellar evolution, significantly affecting stellar lifetimes, yields and luminosities, but we currently lack predictive prescriptions of stellar mass loss. 

Galaxy properties are derived from observations of the integrated light of bright member stars. 
%; Asymptotic giant branch (AGB) stars, 
Stars in the late stages of their evolution are among the infrared-brightest objects in galaxies.  
An unrealistic treatment of the mass-loss process introduces significant uncertainties in
%The mistreatment of the mass-loss process could significantly affect 
galaxy properties derived from their integrated light. 

We describe current efforts and future needs and opportunities to characterize AGB outflows: driving mechanisms, outflow rates, underlying fundamental physical and chemical processes such as dust grain formation, and dependency of these on metallicity.
\end{tcolorbox}

\clearpage

\begin{center}
{\color{orange!50!white}\noindent\hrulefill}{\hspace{0.5cm} \textbf{\Large  Background: stellar outflows }\hspace{0.5cm}}{\color{orange!50!white}\hrulefill\par}
\end{center}
\vspace{-0.4cm}

The large majority of stars experience mass loss via surface outflows over a sizeable fraction of their lives. Low-to-intermediate mass stars  ($0.8$\msun$\lesssim$\mzams$\lesssim 8$\msun) develop increasingly strong outflows during their late evolutionary stages on the asymptotic giant branch (AGB), possibly with episodic modulations. Mass-loss rates are so high ($10^{-8}-10^{-4}$\msun/yr)  that these stars are eventually
stripped of the entire envelopes, leaving white dwarfs as compact remnants. Massive stars  (\mzams$\gtrsim$8\msun) develop strong winds during most of their evolution. Mass loss is expected to significantly affect the structure and evolution at \mzams\,$>15$\msun, as well as the subsequent supernova explosion. Additionally, since the circumstellar environment of a progenitor will be swept up and reprocessed during and after the supernova event\cite{sarangi2018} or possibly even survive \cite{edmunds2005,scicluna2015}, we need to understand the physics and chemistry of these outflows. While the driving mechanisms of mass loss in hot, luminous stars (e.g. blue supergiants, luminous blue variables) are fairly well known, we still lack a comprehensive understanding of the stellar outflows in bright and cool evolved stars (red supergiants and AGB stars), where various complex processes are simultaneously at work: dust formation, stellar pulsation, and convection.

\begin{figure}[h]
\begin{tcolorbox}[breakable,colback=orange!5!white,colframe=orange]
\includegraphics[trim=0 0.3cm 0 0.25cm, clip, width=\linewidth]{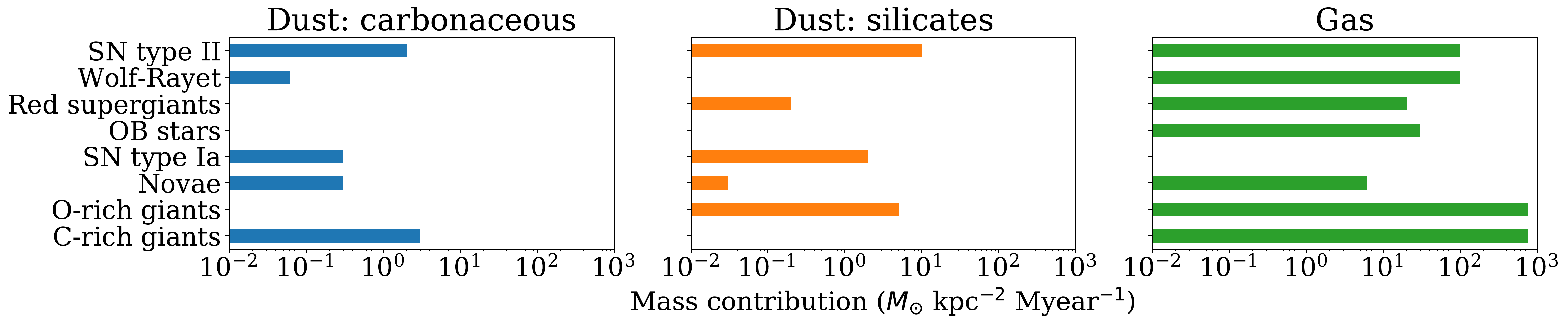}
\vspace{-0.7cm}
\caption{\small Stellar contributions to galactic interstellar gas and dust budgets\cite{tielens2005_book}.
\vspace{-0.3cm}
%: \textit{(left)} carbonaceous dust, \textit{(middle)} silicate and metal dust, \textit{(right)} gas.
 }\label{fig:tielens}
\end{tcolorbox}
\end{figure}

Figure~\ref{fig:tielens} summarizes % the dust and gas 
contributions of stellar outflows and supernovae to the galactic insterstellar medium\cite{tielens2005_book}. %}}
In this paper we address the challenges to characterize outflows from AGB stars in particular. The circumstellar chemistry of AGB stars is set by the atmospheric C/O abundance ratio. The O-rich and C-rich giants listed  in Fig.~\ref{fig:tielens} are AGB stars with ${\rm C/O} <1$ and ${\rm C/O}>1$, respectively. Although the underlying driving mechanisms are different, the outflows from red supergiants and yellow hypergiants, two types of evolved massive stars, are remarkably similar to those of O-rich AGB stars in terms of their gas and dust chemistry. 

%The chemical resemblance between outflows from AGB stars, on the one hand, and RSGs and YHGs, on the other, is remarkable, both for the gas and dust components in these outflows, providing us with an extra test bed for the fundamental gas-dust chemistry discussed before.

\vspace{-0.5cm}
\vspace{-0.55cm}

\vspace{-0.3cm}
\begin{center}
%\textbf{\large I. General summary}\\ \vspace{-0.2cm}
{\color{orange!50!white}\noindent\hrulefill}{ \hspace{0.5cm }\textbf{\Large  Mass-loss rates }\hspace{0.5cm}}{\color{orange!50!white}\hrulefill\par}
\end{center}
%\section{Mass-loss rates}
\vspace{-0.4cm}

\noindent We need to quantify stellar mass-loss rates to constrain the effect of outflows on stellar evolution. By deriving mass-loss rates for both the gas and dust in the outflows ($\dot{M}_{\mathrm{gas}}$, $\dot{M}_{\mathrm{dust}}$) we can obtain a dust-to-gas mass ratio, a proxy for dust-production efficiency. We review the methods used, the uncertainties surrounding these, and the observations needed to advance the field.
\\\vspace{-0.4cm}

%{\color{red} Should the below be reshuffled to first list what we know/don't know and then give the way forward for gas, dust and d/g ratios?}

%\section{Gas-mass-loss rates \label{sect:gasmdot}}  
\noindent\underline{\large\textbf{Gas-mass-loss rates}} of AGB
%, RSG, and YHG 
stars can be derived from observations of CO rotational transitions. Whereas CO\,($J = 1-0$) emission, at 3\,mm wavelength, traces the outermost, coldest molecular gas ($3-10$\,K), higher-excitation transitions ($J > 7-6$) – not observable from the ground – trace the dense, warm ($500-2000$\,K) gas close to the star. A multi-transition approach, covering both the low- and high-excitation emission provides the most reliable outflow models by constraining the temperature and density of the entire circumstellar environment, and with that the mass-loss history of the star. 
Several studies have derived sample statistics or have performed detailed case studies for Galactic sources using ground- and space-based facilities \cite{ bujarrabal2011, ramstedt2008, debeck2010, debeck2012, danilovich2015, nicolaes2018,ramosmedina2018}. In contrast, CO emission has been observed for only 4 extragalactic AGB stars \cite{groenewegen2016}, although it is essential that we understand the effect of metallicity on mass loss.
% {\color{red}Motivate the need for extragalactic objects.} 
 For these four stars, only % the Atacama Large Millimetre/Sub-millimetre Array (ALMA) measured only %one low-excitation transition was traced, 
 the $J=2-1$ emission was measured using the Atacama Large Millimetre/Sub-millimetre Array (ALMA). This emission is not the best outflow tracer since it probes mainly the outermost regions of the outflow, which are
 %severely limiting the information that can be retrieved, since the measured emission traces mainly the outermost regions of the outflow. 
% These are 
 most affected by external factors such as dissociation by interstellar UV irradiation\cite{debeck2010}.
 In addition, this approach can realistically provide $\dot{M}_{\mathrm{gas}}$ for only a limited fraction of the AGB populations in the Small and Large Magellanic Clouds (SMC, LMC), since all CO transitions have to be measured individually.
 
{\color{blue}\bf The way forward:} %An enormous improvement to $\dot{M}_{\mathrm{gas}}$-quantifications can be achieved through
Instantaneous observations of CO transition ladders, covering a large range in excitation properties, can provide an enormous improvement to $\dot{M}_{\mathrm{gas}}$-quantifications. The SAFARI instrument  for ESA's proposed SPICA mission and the OSS instrument  for the proposed {ORIGINS Space Telescope},  broadband direct detectors in the far-infrared (FIR), would provide instantaneous coverage of a large number of CO lines %(SAFARI: $J=12-11,\dots, 79-78$, OSS: $J=5-4, \dots, 104-103$) 
at sensitivities better by over two orders of magnitude than the Herschel Space Observatory/PACS instrument \cite{bradford2018}. The enormous gain in detector sensitivity opens the possibility to measure $\dot{M}_{\mathrm{gas}}$ also for AGB stars in the SMC/LMC, i.e. at significantly lower-than-solar metallicity.  Additionally, {ORIGINS}/OSS would outperform SPICA in these studies with a broader wavelength coverage, a higher spectral resolution (needed to unblend molecular spectra), and a higher spatial resolution (needed to isolate individual extragalactic outflows). Based on Herschel/PACS+SPIRE fluxes for the prototypical source \irc\cite{nicolaes2018}, we could detect CO emission at $\lesssim$300 $\mu$m for on the order 100 evolved stars in the LMC with $\mathrm{S/N}=3-10$ in 1-hour integrations with {ORIGINS}/OSS. This would be limited to stars with $\dot{M}_{\mathrm{gas}}\gtrsim 10^{-6}$\,\msun/yr, meaning that it covers a significant fraction of the AGB population, but misses stars with mass-loss rates $\dot{M}_{\mathrm{gas}}=10^{-8}-10^{-6}$\,\msun/yr. %As CO fluxes roughly scale with $\dot{M}_{\mathrm{gas}}$, it is unlikely that we will observe the CO emission from extragalactic low-$\dot{M}_{\mathrm{gas}}$ stars.
% {\color{red} (These numbers aren't so helpful without describing whether this is typical for AGB stars and/or RSGs/YHGs.)}. %, especially for the oxygen-rich AGB population, where the CO abundance is somewhat lower than in C-rich outflows. %We are not aware of any other existing or planned facilities that can or will be able to perform similar observations.
As we currently do for galactic AGB stars, we could constrain the gas kinetic temperature and density structure of the outflows, and with that possibly the mass-loss history of these extragalactic stars, through radiative-transfer modelling of the CO emission. 
  \\\vspace{-0.35cm}

 \vspace{-0.5cm}
\begin{figure}[h]
\begin{tcolorbox}[breakable,colback=orange!5!white,colframe=orange]
\centering
\includegraphics[trim=0 3.35cm 0 3.6cm, clip, width=0.95\linewidth]{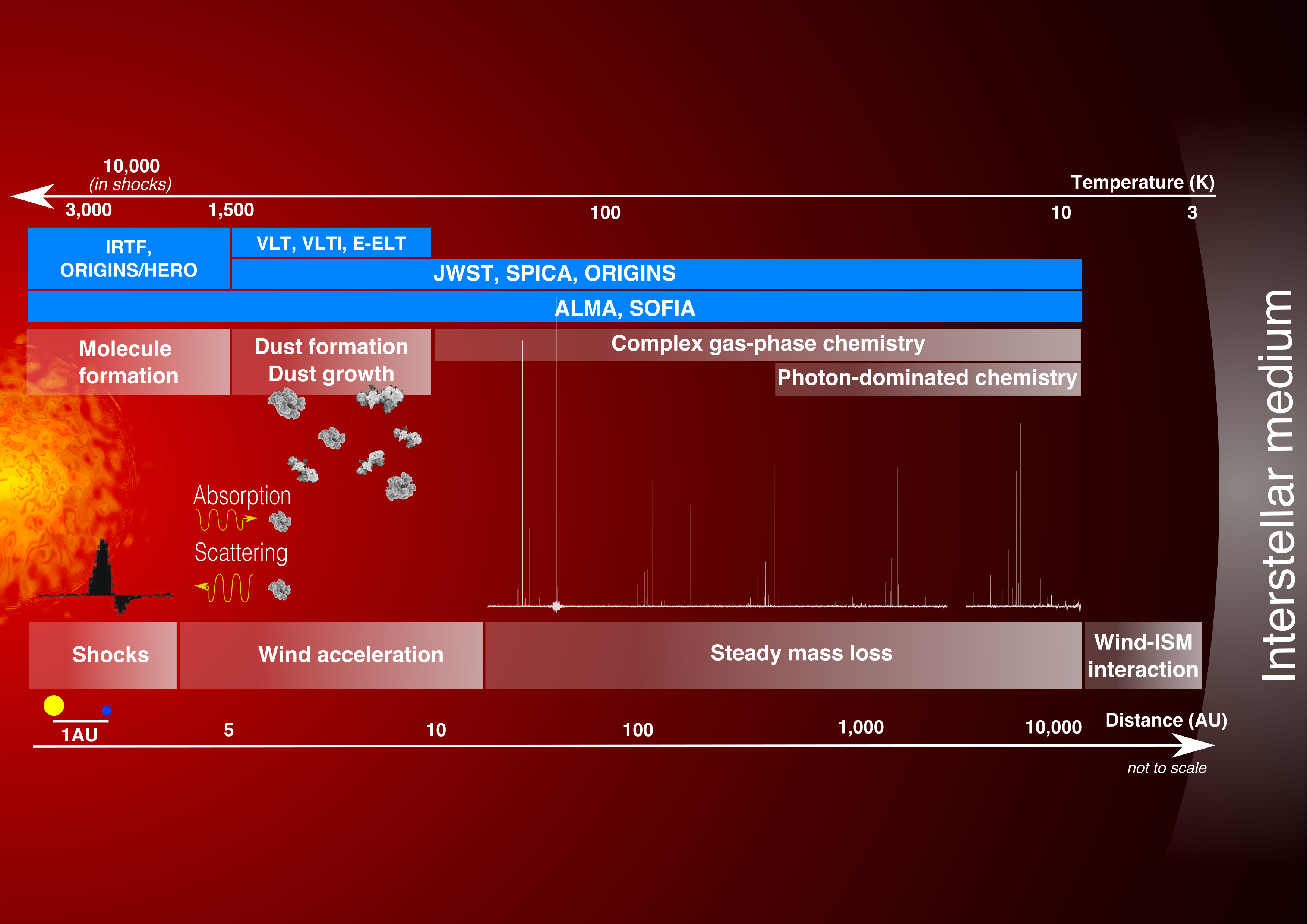}
\caption{AGB outflow structure. \textit{White boxes:} major chemical and physical processes, \textit{blue boxes:} contribution of existing, planned, and proposed facilities.}\label{fig:overview}
\vspace{-0.2cm}
\end{tcolorbox}
\end{figure}  
  \vspace{-0.5cm}
 
\noindent\underline{\large \textbf{Dust-mass-loss rates}} and the bulk properties of circumstellar dust (size distribution, chemical composition, temperature structure) %can be derived from the spectral energy distribution.
%Observations of the spectral energy distribution (SED) of evolved stars provide information on the bulk properties of the circumstellar dust: size distribution, chemical composition, temperature structure, and dust-mass-loss rate. 
%These properties 
%constrain the dust present throughout the outflow and 
are used alongside CO emission to set up empirical outflow models \cite{debeck2012,khouri2014a,khouri2014b,khouri2015,oudmaijer1996}.  
%Different dust species give rise to different opacities (absorption/emission cross-sections), causing significantly different responses to radiation pressure from the incident stellar light. %We discuss the role of these different dust species in \textbf{\color{red}Sect. 4}.
There is a known dichotomy in the chemical properties of the dusty outflows. Amorphous carbon dust and silicon carbide appear together in the outflows of C-enriched stars, whereas %these are stars with an atmospheric carbon abundance exceeding the oxygen abundance 
%(C/O$>$1) {\color{red}(should this maybe be explained already sooner?)}. 
silicates, alumina, and water ice dominate the dust emission of stars with C/O$<$1. 
%{\color{red}MB: This section (or somewhere else) should mention the unknowns in how metallicity affects dust formation.  
The broad spectral features  in the mid-infrared (MIR) of these solids
%; e.g., bands of silicon carbide (SiC) around C-rich stars, % {\color{red}($\lambda\sim XXX$)}, 
%and silicates, %{\color{red}($\lambda\sim XXX$)}, 
%water ice, % {\color{red}($\lambda\sim XXX$)}, 
%and alumina %{\color{red}($\lambda\sim XXX$)} 
%around O-rich stars. These bands which 
have been studied for Galactic sources and in the Local Group, including the Magellanic Clouds, using e.g. UKIRT, IRAS, ISO, Spitzer, and Herschel, over a range of wavelengths and spectral resolving power \cite{kwok1997,mutschke1998,speck1997,speck2000,speck2009,kraemer2002,suh1999,suh2000,suh2002, clement2003,clement2005,messenger2013, boyer2012,boyer2015, riebel2012,srinivasan2016}. Surprisingly, it is found that the Fe-content of AGB dust is extremely low, although iron is significantly depleted from the gas phase in both circumstellar and interstellar environments, and $\sim$90\% of iron produced by stars should be in dust grains \cite{kimura2017}.  Fe-bearing minerals 
%(e.g., melilite, which shows spectral features at $21-24$ $\mu$m) 
are likely on mineralogical grounds but have not yet been found \cite{jones1990_iron,kimura2017}. 

Silicates show multiple emission features in the range $20-100$\,$\mu$m, which depend critically on their temperature and composition, and more specifically, on their relative content of Mg and Fe\cite{devries2012,molster1999,sylvester1999}.  In contrast, metallic iron and amorphous carbon have rather featureless dust continua. Other C-rich dust, like graphite, SiC, and especially polycyclic aromatic hydrocarbons (PAHs) do show emission features at $3-30$\,$\mu$m and the carriers of the 21\,$\mu$m and 30\,$\mu$m features in the outflows of C-rich stars are currently still debated. The feature observed around 30 micron in the outflows of C-stars is often attributed to MgS but its formation is unclear\cite{sloan2014}.

We know very little about the dust composition at low metallicity. Recent studies suggest that carbon dust may form independent of metallicity\cite{boyer2015_metalpoor,boyer2017_highz,goldman2019,bladh2019}, whereas  silicate dust composition likely has a strong metallicity dependence\cite{jones2012}. % in both AGB and RSG outflows.
Moreover, current studies\cite{matsuura2009,matsuura2013} suggest that the fraction of C-rich AGB stars is higher at lower metallicity, and, conversely, that the fraction of O-rich AGB stars decreases with lower metallicity. However, there is a possibility that the current surveys miss higher-$\dot{M}_{\mathrm{gas}}$ O-rich AGB stars that could significantly contribute to the interstellar enrichment.
%These results need to be confirmed with future observatories such as JWST and {ORIGINS}. Especially, {ORIGINS} and SPICA will provide further details on the chemical composition with its much broader wavelength coverage. {\color{red} add sth about large aperture?} %{\color{red}(about o-rich, much broader wavelength range needed) However,  with current samples where we can detect this 50 micron regions for stars is currently limited to the ISO/SWS sensitivity range.}
%though the DUSTiNGS survey (Boyer et al. 2015b \cite{boyer2015_metalpoor} and Boyer et al. 2017 \cite{boyer2017_highz} and Goldman et al. 2019 \cite{goldman2019}) hints that at least carbon dust may form independent of metallicity.  
%
%Dust budgets have been measured for both galactic\cite{xxx} {\color{red}(REFS)} and extragalactic AGB stars \cite{boyer2012,boyer2015, riebel2012,srinivasan2016} using {\bf\color{red} INSTRUMENTS (ISO, IRAS, Spitzer,Herschel ?? - i.e. wavelength coverage, anything obvious missing?)}.  
Additionally, the balance between AGB stars and supernovae as sources of dust input to the interstellar medium remains heavily debated and it is thus crucial to characterize AGB populations and quantify their dust-mass loss at different metallicities. % and to characterize the composition of the dust in these outflows. 

{\color{blue}\bf The way forward:} 
%, iron bearing crystalline silicates show a strong dependence on wavelength\cite{molster1999,sylvester1999}.  
%{\color{red}  Add sth in particular about carbon dust as well? MgS ?}
%{\color{red}metallicity dependence: c not, si yes} These results need to be confirmed with future observatories such as JWST and {ORIGINS}. Especially, {ORIGINS} and SPICA will provide further details on the chemical composition with its much broader wavelength coverage. {\color{red} add sth about large aperture?}
The dust composition and $\dot{M}_{\mathrm{dust}}$ of outflows in the Galactic Bulge and throughout the Local Group (at much lower-than-solar metallicities) will be investigated with the MIRI and NIRCam %{\color{red}(goes to 5 microns)} 
instruments on board the James Webb Space Telescope (JWST), but {ORIGINS} and SPICA will provide further details on the chemical composition with their much broader wavelength coverage. These missions could observe the outflows of additional Galactic AGB stars which are not feasible with JWST because of saturation, and of AGB stars in the Local Group that can be spatially isolated.
%SPICA and  { ORIGINS} could measure the dusty component in the outflows of additional Galactic AGB stars, likely not feasible with JWST because of saturation, and of AGB stars in the Local Group that can be spatially isolated. 
%{\color{red}more details needed here on added wavelength points, sensitivity, resolution, variability etc.?}  %{\color{blue}Especially, {ORIGINS} and SPICA will provide further details on the chemical composition with their much broader wavelength coverage. }
SOFIA/FORCAST can be used for Galactic sources. % at resolutions $R\sim100-300$ in the range $5-40$\,$\mu$m.
 %{\color{red}Other obvious features/species to mention here to cover other wavelengths? } For Galactic sources, we can target part of this range using SOFIA/FORCAST, which provides us with a $R\sim100-300$ in the range $5-40$\,$\mu$m. For extragalactic sources, JWST/MIRI, SPICA, and {ORIGINS}/MISC and ORIGINS/OSS will be needed and are complementary in their wavelength coverage. 
 % You can mention also SPICA/SMI here as this Japanesse instrument will work in range 12-36 microns and will be suitable for dust features in this range, especially for low resolution mode R=100~20000 in thw range 17-36 microns.
 \\\vspace{-0.35cm}
 
\noindent\underline{\large\textbf{Dust-production efficiency}} can be derived for a large part of the Galactic AGB population as we have estimates of both $\dot{M}_{\mathrm{gas}}$ and $\dot{M}_{\mathrm{dust}}$. %, {\color{red}although there are uncertainties}\cite{lombaert2013}. %depending on the method of derivation\cite{lombaert2013}. %, that become increasingly reliable because of improved distances delivered by e.g. Gaia.
We do not have reliable dust-to-gas ratios for extragalactic AGB stars since we have virtually no $\dot{M}_{\mathrm{gas}}$-estimates\cite{groenewegen2016} (see above) and we instead have to rely on indirect $\dot{M}_{\mathrm{gas}}$-measurements\cite{vanloon2000}.
 %. Reliable dust budgets have been measured for extragalactic stars \cite{boyer2012,boyer2015, riebel2012,srinivasan2016}, but only for 4 extragalactic AGB stars have observations of CO emission been obtained \cite{groenewegen2016}. 
 To genuinely translate dust-mass-loss rates into a degree of chemical enrichment of the host galaxies, we need to understand how efficiently the dust is produced. Constraints on the dust-production efficiency in environments of substantially different metallicity are especially important in  light of the discovery of extreme AGB stars producing about 65\% of the total dust budget in the Magellanic Clouds and M32, but accounting for only 3\% of the AGB population\cite{jones2015,boyer2012,riebel2012,srinivasan2016}. Are these objects also extreme in their gas-mass-loss rates or is their dust-production efficiency substantially higher? How does this relate to the metallicity? %{\color{red}MB: This is an interesting question, but you don't really come back to it or describe explicitly how you'll answer this question.}
The proposed measurements of $\dot{M}_{\mathrm{gas}}$ and $\dot{M}_{\mathrm{dust}}$, outlined above, will directly serve to answer these questions.
\vspace{-0.3cm}
\begin{center}
%\textbf{\large I. General summary}\\ \vspace{-0.2cm}
{\color{orange!50!white}\noindent\hrulefill}{ \hspace{0.5cm }\textbf{\Large  Outflow drivers }\hspace{0.5cm}}{\color{orange!50!white}\hrulefill\par}
\end{center}
 \vspace{-0.4cm}

\noindent\underline{\large\bf Shocks} 
Large-scale convective motions inside an AGB star replenish the outer stellar layers with newly processed elements and cause granulation at the surface \cite{freytag2017,paladini2018}. In combination with long-period (100-1000\,days) pulsations, these motions cause shocks in the star’s upper atmosphere, which can manifest themselves as a low-filling factor chromosphere with temperatures up to a few $10^4$\,K \cite{vlemmings2017}. For reference, the effective temperature of an AGB star is typically in the range $2000-3000$\,K\cite{debeck2010}. The high temperatures and densities in the shocked material cause an out-of-equilibrium chemistry %that leads to a significantly different {\color{red}(from what?)}
%in the molecular and possibly also in the dusty content of 
%in the outflow 
\cite{cherchneff2006,cherchneff2012,gobrecht2016,marigo2016}. The main questions surrounding shocks and their role in outflow driving are the following:  How large is the region affected by shocks? Which densities and temperatures are reached in the shocked material? How do the shocks relate to the pulsation of the star, both in periodicity and strength? 
 
{\textbf{ \color{blue} The way forward: }} Observations of the stellar continuum and of molecular tracers, such as SiO or vibrationally excited CO, can reveal the size of the pockets of shocked material\cite{khouri2016_co,vlemmings2017,wong2016} and provide constraints on the temperature and density of the shocked gas. Near-infrared observations with e.g. IRTF/iSHELL of molecular bands of species that are highly abundant in the upper stellar atmosphere -- such as C$_2$H$_2$, HCN, SiO, or H$_2$O -- could trace the chemical effect of shocks; IRTF/TEXES and SOFIA/EXES could do this for matter at a few to a few tens of stellar radii; whereas SOFIA/GREAT and {ORIGINS}/HERO could do this in the far-infrared.

To disentangle the complex velocity field at the base of the outflow with infall and outflow, we require a resolution better than 1\,km/s, since the contributions of molecular absorption and emission otherwise become indistinguishable or undetectable.  Moreover, the variability in the emitted light over time can only be characterized using multi-epoch observations. Multi-epoch, multi-frequency observations with ALMA can already now trace shock waves as they propagate through the atmosphere in submm/mm tracers \cite{kaminski2017,khouri2019}. 
%{\color{blue}The disturbance of equilibrium chemistry provides the possibility to trace shocks by probing the near-infrared of molecular bands of species that are highly abundant in the upper stellar atmosphere -- such as C$_2$H$_2$, HCN, or H$_2$O. This would be possible for Galactic targets using e.g. IRTF/iSHELL for the upper atmosphere of the star, and IRTF/TEXES and SOFIA/EXES for layers from a few to a few tens of stellar radii. In order to disentangle effects of molecular excitation and chemistry, we would need similar observations in the FIR which will only be possible with SOFIA/GREAT and {ORIGINS}/HERO, for low-intensity signatures. Both sets of observations would be limited to Galactic sources.}
%{\color{red}JPF: High spectral resolution observations in the near-IR could also help a lot. You can check the papers by Hinkle on R Leo (Hinkle, 1978, ApJ, 220, 210; Hinkle \& Barnes, 1979, ApJ, 227, 923). Now, this could be done with IRTF/iSHELL, for instance. This for the surroundings of the star. For the region from 3-5 to 30R* (depending on the star), IRTF/TEXES would be very useful. SOFIA/EXES as well but I think it will be decommissioned. Of course, all this for Galactic targets.}
The in-depth treatment of the molecular excitation using detailed, non-LTE radiative transfer models will be an essential step in the analysis to characterize the effect of the variations in the stellar radiation field on the line emission\cite{cernicharo2014,pardo2018}, as opposed to changes in the molecular abundance caused by shocks.
%{\color{blue}As we described for the analysis of shocks, radiative transfer models will be essential to disentangle effects of variability in the stellar radiation, in the chemistry, and in the excitation.}
\\\vspace{-0.35cm}

\noindent\underline{\large \bf Dust } forms in the warm regions (1000-1500\,K) close to the AGB  star, although observations suggest that the smallest dust grains could form at even higher temperatures \cite{khouri2018}. The grains experience radiation pressure through absorption or scattering of incident stellar light and are pushed radially outward. 
%The high densities in these regions ensure momentum transfer from the dusty to the gaseous component, leading to steady outflows that build large circumstellar envelopes of gas and dust \cite{hoefner2018}. 
Momentum transfer to the gas leads to steady outflows that build large circumstellar envelopes of gas and dust \cite{hoefner2018}. 
%The outflows are not necessarily smooth, in both space and time, as is seen both from observations \cite{mauron2000} and from the coupled hydrodynamical treatment of the two fluid components \cite{simis2001}.
%
%The current paradigm states that dust grains close to the AGB star experience radiation pressure through absorption or scattering of incident stellar light and are, consequently, pushed radially outward. The high densities in these regions ensure momentum transfer from the dusty to the gaseous component, leading to steady outflows that build large circumstellar envelopes of gas and dust \cite{hoefner2018}. 
%
%\underline{\bf 1. Shocks} caused by stellar pulsations and large-scale convective flows produce severely increased temperatures and densities in the stellar-surface layers over short timescales {\color{red}(REFERENCE)}. 
%
Since the opacities of different dust species cause significantly different responses to radiation pressure from the incident stellar light, wind-driving models depend critically on the dust properties. Grain size, density, and opacity (absorption/emission cross-section) need to be well-constrained, especially at the base of the outflows\cite{hoefner2008}.  The observed outflow rates in the case of C-rich AGB stars can be reproduced, given optical properties of the carbonaceous dust. For O-rich AGB stars, however, a number of open questions remain. %the observed outflow rates cannot easily be reproduced. 
Alumina form closest to the star, at $\sim$2\rstar \cite{karovicova2013, khouri2015, khouri2018, zhaogeisler2012} but cannot drive the wind \cite{hoefner2016}.  Iron-bearing silicates seem to represent a larger mass fraction of the total dust in the outflow (e.g. $\sim$65\% in the case of W Hya\cite{khouri2015}), but are found only beyond 5\rstar \cite{zhaogeisler2012} which rules them out as drivers at the base of the outflow. 
In order for radiation pressure to be sufficient, scattering of stellar light on large Fe-free silicate grains close to the star has been suggested\cite{hoefner2008}. The presence of large grains has been corroborated\cite{norris2012,ohnaka2016,ohnaka2017}, but the micro-physics of their formation processes remains unclear. Which seed particles do the grains grow on? How efficient is the gradual Fe-enrichment of silicates in the wind?

{\textbf{ \color{blue} The way forward: }}
%The  composition, size, location, and density distribution of the dust are essential parameters in the wind driving mechanism\cite{hoefner2008}. 
Observations at high-angular resolution using e.g. VLT/SPHERE, VLTI/MATISSE and the future E-ELT/METIS can provide the spatial information, size, and density of dust close to the star \cite{khouri2016,khouri2018,ohnaka2016,ohnaka2017} in the case of Galactic sources.  If {ORIGINS}/MISC carries a coronagraph, it could trace the emission from solid-state bands at distances of a few to a few tens of stellar radii in these outflows, where the dust grains grow to their final size and composition.
%These results can be used also to interpret the dust signature of extragalactic sources, where this spatial information is not accessible.
Improved knowledge of the bulk dust properties (see above) will set strong constraints on the possible dust-condensation sequences.

\vspace{-0.2cm}
\begin{center}
%\textbf{\large I. General summary}\\ \vspace{-0.2cm}
{\color{orange!50!white}\noindent\hrulefill}{ \hspace{0.5cm }\textbf{\Large  Dust formation }\hspace{0.5cm}}{\color{orange!50!white}\hrulefill\par}
\end{center}
\vspace{-0.4cm}

\noindent Large uncertainties in the current wind-driving paradigm relate to the path of dust formation and growth. Which molecules form the first clusters that serve as seeds for solid-grain growth? Which are the relevant condensation sequences? What are the properties (size, shape, opacity) of the grains close enough to the star to drive a wind? Do these vary over the stellar pulsation cycle? If so, how does this affect the outflow? Do these properties vary with metallicity?
%\\\vspace{-0.3cm}

%\section{The problems of dust formation}
\noindent\underline{\large\textbf{Gas depletion}} and the chemical processes involved in the transition between the gas and solid phases around AGB stars of all chemical types are currently only poorly characterized. Several theoretical efforts have recently been made to understand this critical part of circumstellar chemistry \cite{boulangier2019, gobrecht2016,gobrecht2017}. Empirical constraints have to come from observations of both the dusty and gaseous components in the upper atmosphere and at the base of the outflow of all chemical types of AGB stars \cite{cernicharo2015, debeck2015,debeck2017,decin2017,fonfria2008,kaminski2013,kaminski2016, khouri2016,khouri2018,massalkhi2018}. Additionally, there is a lack of characterization (both observationally and theoretically) of the effect of dust grains on the gas chemistry. What is the role of grain-surface reactions, evaporation, sputtering, etc.?%We outlined above that the size, density, and chemical composition of dust close to the star can be constrained with current and upcoming facilities. 
%(e.g. VLT/SPHERE, VLTI/MATISSE, E-ELT/METIS). 
%{\color{red}very short on ALMA results}

{\bf \color{blue} The way forward:} Current efforts to measure the depletion of gas-phase species in the dust-formation and growth processes lack access to sensitive, high-spectral resolution observations in the far-infrared. Whereas ALMA can observe at high-angular resolution and extremely high sensitivity, a high-spectral resolution instrument on board a space-based facility like {ORIGINS}/HERO  is needed to access transitions of H$_2$O and hydrides, as well as high-excitation lines of numerous other molecules considered critical in the dust formation and growth (e.g. Ti-, Al-, Fe- bearing molecules)\cite[][]{kaminski2016,kaminski2017}. No other planned facility will be able to address this aspect of the fundamental question of how dust is formed.
High-spatial resolution observations with e.g. ALMA are additionally needed to study the effect of dust grains on gas chemistry\cite{decin2016}.
%{\color{red}Extra gas chemistry through grain-surface reactions, evaporation, sputtering, etc?}
\\\vspace{-0.35cm}

\noindent\underline{\large\bf The first solids} to appear in the outflows have not yet been characterized. In the case of O-rich stars, large gas-phase clusters such as (Al$_2$O$_3$)$_n$ are likely candidates to form the first dust grains. For C-rich stars, PAHs, have been suggested\cite{cherchneff2000,cherchneff2012}. 

{\bf \color{blue} The way forward:} Clusters and PAHs show broad spectral features in the mid-/far-infrared much like most dust species\cite{decin2017,draine2011} and there is a need for spectroscopic predictions. The enormous wavelength coverage of  SPICA and ORIGINS/OSS would be ideal to search for these.
E-ELT/METIS observations will be essential to test the presence of these very small clusters.
\\\vspace{-0.35cm}

\noindent\underline{\large\bf Variability: } Ideally, future observations of gas depletion and dust formation will sample the gas and dust (quasi-)simultaneously, to mitigate  effects of variability on the analysis, and at multiple epochs in time\cite{khouri2018}, to characterize the chemical variability in the material throughout the pulsation cycles of the stars  (periods of a few hundred days) and provide empirical constraints for true time-domain astrochemistry. 
%{\color{red}The in-depth treatment of the molecular excitation using detailed, non-LTE radiative transfer models will be an essential step in the analysis to mitigate the effect of the variations in the stellar radiation field on the line emission\cite{cernicharo2014,pardo2018}, as opposed to changes in the molecular abundance caused by shocks.}
%Observations of these signatures are ideally coupled to the observations that measure the gas depletion .
%Although the spectral resolving power in  is very low {\color{red}(XX)} 
The transit channel of {ORIGINS}/MISC could measure the temporal variations in the dust emission in the outflow (see also above). Although it is optimized for high-cadence time series (5s--1min), we could get low-spectral resolution variability measures on timescales of tens of days. 

 \vspace{-0.3cm}
\begin{center}
{\color{orange!50!white}\noindent\hrulefill}{\hspace{0.5cm} \textbf{\large  Prospects and needs }\hspace{0.5cm}}{\color{orange!50!white}\hrulefill\par}
\end{center}
 \vspace{-0.3cm}

 %{\color{red}This section needs a lot of work.  As it is, it doesn't really tie back to what you've written.  And it introduces two new topics (theory and hydrodynamical models), which are suddenly introduced with no real context in how it applies to everything you've discussed.}

\noindent\underline{\textbf{ Observations}} from current and upcoming facilities need to provide strong constraints on dust formation and growth. This includes ground- and space-based facilities covering an enormous wavelength range and covering a large variety of observational techniques.
%\\\vspace{-0.3cm}

\noindent\underline{\textbf{ Theory:}} The spectroscopy of molecules and clusters needs to be expanded in order to successfully search for and identify the relevant spectral features.  More theoretical efforts are also needed to describe the chemical processes involved in the dust-gas chemistry, such as molecular-cluster formation, grain-surface chemistry, and evaporation of gas from the grain surfaces. To consider the effects of large-scale convective flows on the wind driving, 3D radiation-hydrodynamical models of dust-driven winds need to be developed\cite{hoefner2019}, to complement the spherically symmetric models used at present\cite{mattsson2010,eriksson2014,bladh2015,bladh2019}.  In addition, there will be a need for hydrodynamical models that study the influence of the interaction between the stellar outflows and the ISM on the properties of the dust grains that are deposited into the ISM. Finally, results from wind models  and observations (mass-loss rates, molecular abundances, dust-to-gas ratios)  may be included in stellar evolution calculations for evolved stars\cite{marigo2013,marigo2016}  to test their impact as a function of stellar mass and metallicity.

\clearpage
\addcontentsline{toc}{chapter}{Bibliography}
\bibliographystyle{aa}
\setlength{\bibsep}{0.0pt}
\small
\begin{multicols}{2}
\bibliography{WhitePaper.bib}
\end{multicols}
\normalsize

%
%
%
%\newpage
%
%
%
%{\color{red} MB: The first and last paragraphs must be re-written. It is not at all clear from either what this white paper is about. In particular, the first paragraph spends a lot of time talking about stellar mass, which ultimately isn’t all that important to understanding our goals here.  And in the last paragraph, theory and models are suddenly introduced without much context at all, while the observations (which are the main thrust of this paper) are mostly ignored.
%\begin{itemize}
%\item a summary paragraph either at the beginning or end describing which missions you're addressing in this white paper and which questions you'll answer with those missions.
%\item a summary table with the following columns: (a) Mission (b) Instrument (c) Capabilities (like wavelength coverage, resolution, etc.), (d) Measurables (e.g., like dust composition, gas-to-dust ratio), and (e) Key outflow questions addressed.  A table like this has the added benefit that you can continually refer to it in the text so you don't have to repeat yourself.
%\end{itemize}
%
%Maybe restructure as:
%\begin{enumerate}
%\item Background (more or less your section 1)
%\item Current State of Affairs (i.e., what we know)
%\item Key Outstanding Questions (especially if you can sum these up into bullets)
%\item Future Prospects/Needs (i.e., what we need to answer those questions, both observationally and with theory, etc.).
%\end{enumerate}
%
%}

\end{document}